\documentclass[12pt]{article}
\usepackage{cite,epsfig,amssymb,amsmath,graphicx,color}
\usepackage{cite,epsfig,amssymb,amsmath,graphicx,color}
\newcommand{\nn}{\nonumber}

\newcommand{\1}{1\kern -3pt \mathrm{l}}

\begin{document}

\begin{center}
{{{\Large \bf Spontaneous Supersymmetry Breaking in Inhomogeneous Supersymmetric Field Theories and BPS Vacua }
}\\[17mm]
~~Yoonbai Kim$^{1}$,~~O-Kab Kwon$^{1}$,~~D.~D. Tolla$^{1,2}$\\[3mm]
{\it $^{1}$Department of Physics,~BK21 Physics Research Division,
 Autonomous Institute of Natural Science,~Institute of Basic Science, Sungkyunkwan University, Suwon 16419, Korea\\
$^{2}$University College,
Sungkyunkwan University, Suwon 16419, Korea}\\[2mm]
{\it ~yoonbai@skku.edu,~okab@skku.edu,~ddtolla@skku.edu}}

\end{center}
\vspace{15mm}

\begin{abstract}
We study spontaneous supersymmetry breaking in  inhomogeneous extensions of ${\cal N}=1$ supersymmetric field theory models in 4-dimensions. The ${\cal N}=1$  Abelian Higgs model with the inhomogeneous mass parameter and the FI coefficient that are  dependent on spatial coordinates, as well as the O'Raifeartaigh model with all its parameters being  dependent on spatial coordinates, are studied in detail. In the presence of inhomogeneous parameters, half supersymmetry can be preserved by adding appropriate inhomogeneous deformations to the original  Lagrangians. The inhomogeneous deformations often break the R-symmetry explicitly. In cases where the inhomogeneous deformations do not break the R-symmetry explicitly, we demonstrate that spontaneous breaking of the R-symmetry is infeasible. We argue that those models can not be spontaneous supersymmetry breaking models, according to the Nelson-Seiberg argument. We comment on this issue  in the context of a generic ${\cal N}=1$ supersymmetric model as well.
\end{abstract}

\newpage
\tableofcontents

\section{Introduction}
Field theory models in which all or some of their parameters depend on spatial coordinates are referred to as inhomogeneous models.  They caught some attention in the investigations of gauge-gravity duality  as such inhomogeneous field theories are dual to gravity theories with spatially varying  backgrounds on the boundary of AdS spacetime~\cite{Bak:2003jk, Clark:2005te, DHoker:2006qeo, DHoker:2007zhm, Kim:2008dj, Kim:2009wv, Gaiotto:2008sd, Hashimoto:2014vpa, Choi:2017kxf, Kim:2018qle, Kim:2019kns, Anderson:2019nlc, Arav:2020obl, Kim:2020jrs}. The story started with the dual field theory to a space-dependent dilaton solution of the type IIB supergravity theory, which is the 4-dimensional supersymmetric Yang-Mills  theory with a gauge coupling that varies across a co-dimension one interface, and approaches two distinct constant values in the limit of the two asymptotic 
boundaries~\cite{Bak:2003jk, Clark:2005te, DHoker:2006qeo, DHoker:2007zhm, Kim:2008dj, Kim:2009wv, Gaiotto:2008sd, Hashimoto:2014vpa, Choi:2017kxf}. Following this, investigations were also extended to the field theories with inhomogeneous mass parameters~\cite{Kim:2018qle, Kim:2019kns, Anderson:2019nlc, Arav:2020obl, Kim:2020jrs}, which were identified as dual to supergravity solutions on the backgrounds of inhomogeneous field strengths~\cite{Gauntlett:2018vhk, Arav:2018njv, Ahn:2019pqy, Hyun:2019juj, Arav:2020asu, Dedushenko:2021mds, Chen:2021mtn}. 
The study of such inhomogeneous field-theoretic models is particularly relevant in the description of defects in two and three dimensional field theories~\cite{Hook:2013yda, Tong:2013iqa, Adam:2019yst, Adam:2018pvd, Adam:2018tnv, Adam:2019djg, Manton:2019xiq, Adam:2019hef, Adam:2019xuc},\footnote{ See also \cite{Kwon:2021flc, Kwon:2022fhv, Ho:2022omx} for a systematic study on constructing inhomogeneous supersymmetric field theory models in two dimensions.} which  play an important role in examining interesting phenomena in diverse areas of physics, ranging from condensed matter  to particle theory and cosmology.

Supersymmetry plays a crucial role in rendering field theory models solvable, but to make connections with the real world physics, it is important to understand how supersymmetry can be broken. However, the mechanisms of supersymmetry breaking in inhomogeneous supersymmetric field theory models have not been explored yet. In this paper, we investigate spontaneous supersymmetry breaking in a selection of inhomogeneous ${\cal N}=1$ supersymmetric theories in (1+3)-dimensions. The Nelson-Seiberg argument~\cite{Nelson:1993nf} imposes stringent constraints on the pattern of spontaneous supersymmetry breaking. According to this argument, the existence of an R-symmetry is a necessary condition for spontaneous supersymmetry breaking, and the presence of a spontaneously broken R-symmetry is the sufficient condition. Inhomogeneous deformations of supersymmetric field theories often explicitly break the R-symmetry. Therefore, to have any luck of building an inhomogeneous model of spontaneous supersymmetry breaking, one needs to find a way to introduce the inhomogeneity without explicit breaking of the R-symmetry.

In our study, we mainly focus on 4-dimensional inhomogeneous supersymmetric Abelian Higgs model and an inhomogeneous extension of the O'Raifeartaigh model~\cite{ORaifeartaigh:1975nky}, and explore the spontaneous supersymmetry breaking. In the case of the O'Raifeartaigh model, we find that the inhomogeneous deformations explicitly break the R-symmetry, and outright making it impossible to construct an inhomogeneous model of spontaneous supersymmetry breaking. In the Abelian Higgs model, we consider two scenarios. First, we examine the case of inhomogeneous mass parameter, which breaks the R-symmetry. Second, we investigate the case of an inhomogeneous Fayet-Iliopoulos (FI) coefficient~\cite{Fayet:1974jb}, which preserves the R-symmetry. By virtue of the Nelson-Seiberg argument~\cite{Nelson:1993nf}, spontaneous supersymmetry breaking could be realized in the latter case, and we discuss the possibility for such breaking to or not to occur. We briefly discuss this issue in a generic field theory model as well.

The remaining part of the paper is organized as follows. Section 2 presents the construction of the inhomogeneous supersymmetric Abelian Higgs model. We describe it initially in the component field form, and then in the ${\cal N}=1$ superfield formalism.
In Section 3, we derive the Bogomol'nyi-Prasad-Sommerfield (BPS) equations for the inhomogeneous supersymmetric Abelian Higgs model. Drawing upon the Nelson-Seiberg argument, we discuss the existence of vacuum solutions and explore the possibilities for spontaneous supersymmetry breaking.
In Section 4, we aim to generalize our findings by examining the spontaneous supersymmetry breaking in the inhomogeneous extension of the O'Raifeartaigh model, once again utilizing the Nelson-Seiberg argument. We also provide a brief overview of these aspects in some generic models.
Finally, in Section 5, we summarize our findings and draw some conclusions.

\section{ Inhomogeneous Supersymmetric  Abelian Higgs Model}

Though the homogeneous supersymmetric models are obtained as an extension of the Poincare  symmetry, inhomogeneous supersymmetric models, which preserve  Poincare symmetry partially,  were also investigated in various models. In these inhomogeneous models, supersymmetry can be partially preserved by an appropriate choice of the spatial dependence of the parameters of the theory. In this section, we consider  supersymmetric Abelian Higgs model in which the mass parameter and the FI coefficient are  space-dependent.
\subsection{Component field formalism}
Lets start by summarizing the massive 4-dimensional $ {\cal N}=1$ homogeneous supersymmetric Abelian Higgs model using the component field formalism. The field content of this model consists of
one massless vector multiplet $(A_\mu,\lambda)$ and two oppositely charged massive chiral multiplets $(\phi_a,\psi_a)$ with $a=1,2$. 
Here, the gaugino field $\lambda$ is a two component Majorana spinor, whereas the two fermionic fields $\psi_a$ in the chiral multiplets are two component Weyl spinors.  The Lagrangian of the  model is  
\begin{align}\label{L-massive}
{\cal L_{\rm SAH}}=&-\frac1{2} F_{\mu\nu}F^{\mu\nu}-i\bar\lambda\bar\sigma^\mu \partial_\mu\lambda-D_\mu\bar\phi_a D^\mu\phi_a-i\bar\psi_a\bar\sigma^\mu D_\mu\psi_a-m^2\big(|\phi_1|^2+|\phi_2|^2\big)
\nonumber\\
&+im(\psi_1\psi_2-\bar\psi_1\bar\psi_2)-ig\Big[\lambda(\psi_1\bar\phi_1-\psi_2\bar\phi_2)-\bar\lambda(\bar\psi_1\phi_1-\bar\psi_2\phi_2)\Big]
\nonumber\\
&-\frac{g^2}4 \big(|\phi_1|^2-|\phi_2|^2-\xi\big)^2.\end{align}
Before inhomogeneity is introduced, the mass parameter $m$, the gauge coupling $g$, as well as  the FI coefficient $\xi$  are constants~\cite{Wess:1974jb}.   The $2\times2$ $\sigma$-matrices, $\sigma^\mu$ and $\bar\sigma^\mu$,  are composed of the Pauli matrices $\sigma^i$ and the identity matrix ${\rm I}$ as 
\begin{align}
&\sigma^\mu=({\rm I},\sigma^i),\qquad\bar\sigma^\mu=({\rm I},-\sigma^i),\nn
\end{align}
and subsequently
\begin{align}
&\sigma^{\mu\nu}=\frac14(\sigma^\mu\bar\sigma^\nu-\sigma^\nu\bar\sigma^\mu),\qquad \bar\sigma^{\mu\nu}=\frac14(\bar\sigma^\mu\sigma^\nu-\bar\sigma^\nu\sigma^\mu).
\end{align}
The bars over the fermionic and bosonic fields denote Hermitian conjugates throughout this work. The covariant derivatives $D_\mu$ for the two charged complex scalar fields and the two charged complex  spinors are given by \begin{align}
D_\mu \begin{pmatrix}
\phi_a \\ \psi_a \end{pmatrix} = \big[\partial_{\mu} -(-1)^a ig A_\mu \big] \begin{pmatrix} \phi_a \\ \psi_a \end{pmatrix}.
\end{align}

The Lagrangian  
\eqref{L-massive} is invariant under the supersymmetry transformation:
\begin{align}\label{delta-2}
&\delta_\epsilon\phi_a=\epsilon\psi_a, \qquad \delta_\epsilon\bar\phi_a= \bar\psi_a\bar\epsilon,\nn\\
&\delta_\epsilon\psi_a= i\sigma^\mu\bar\epsilon D_\mu\phi_a -im_{ab} \epsilon\bar\phi_b,\qquad \delta_\epsilon\bar\psi_a =i\bar\sigma^\mu\epsilon D_\mu\bar\phi_a +im_{ab} \bar\epsilon\phi_b,\nn\\
&\delta_\epsilon\lambda=-\sigma^{\mu\nu}\epsilon F_{\mu\nu}+\frac12ig\big(|\phi_1|^2-|\phi_2|^2-\xi\big)\epsilon,\qquad \delta_\epsilon\bar\lambda=\bar\epsilon \bar\sigma^{\mu\nu}F_{\mu\nu}-\frac12ig\big(|\phi_1|^2-| \phi_2|^2-\xi\big)\bar\epsilon,\nn\\
&\delta_\epsilon A_\mu=-\frac {i}2(\bar\epsilon\bar\sigma_\mu\lambda-\bar\lambda\bar\sigma_\mu\epsilon),
\end{align} 
where the $2\times 2$ mass matrix is symmetric and off-diagonal $m_{ab} = \begin{pmatrix}
0 & m\\
m &0
\end{pmatrix}$, and the supersymmetry parameter $\epsilon$ is a two component constant Weyl spinor.

 Now, lets introduce inhomogeneity to the model of our interest \eqref{L-massive}  by taking into account  the mass parameter $m$ and the FI coefficient $\xi$ as functions of spatial coordinates $x^i$. Then, the supersymmetric variation of the Lagrangian density \eqref{L-massive} generates such terms involving the derivatives of mass and FI coefficient
 \begin{align}\label{deltaL3-2g}
\delta_\epsilon {\cal L}_{\rm SAH}&= (\partial_\mu m_{ab})\big(\psi_a \sigma^\mu\bar \epsilon\phi_b+\epsilon\sigma^\mu\bar\psi_a \bar\phi_b\big)-\frac g2 \big(\partial_\mu\xi\big)(\bar\epsilon\bar\sigma^{\mu}\lambda+\bar\lambda\bar\sigma^{\mu}\epsilon).
\end{align}
In order to achieve an inhomogeneous supersymmetric massive  Abelian-Higgs model which preserves partial supersymmetry, there are two plausible ways of our interest: One is to assume $x^1$-dependent mass function $m(x^1)$ while keeping the FI coefficient to be constant. The other is to relax the FI coefficient to be a function of two spatial coordinates $\xi(x^1, x^2)$ but keep the mass parameter constant. 

We explore first the case that  the mass  $m$  depends on one spatial coordinate $x^1$ and  the FI coefficient $\xi$ is a nonzero constant
\begin{align}\label{mux}
m=m(x^1),\qquad \xi \ne 0~~{\rm with}~~ \partial_{\mu} \xi = 0. 
\end{align}
If we choose and impose the  following projection on the supersymmetry parameter,
\begin{align} \label{Sig-epsilon}
\sigma^1\bar\epsilon=\epsilon~ \Longleftrightarrow~ \epsilon\sigma^1=\bar\epsilon
\end{align}
which reduces the number of supersymmetry by half, then  the second term of the supersymmetric variation \eqref{deltaL3-2g} becomes zero whereas the first becomes
 \begin{align}\label{deltaL3-2g-2}
\delta_\epsilon {\cal L}_{\rm SAH}&= m'\big(\psi_1\epsilon\phi_2+\psi_2\epsilon\phi_1+\bar\epsilon\bar\psi_1\bar\phi_2+\bar\epsilon\bar\psi_2\bar\phi_1\big),
\end{align}
where $m' \equiv \frac{dm}{dx^1}$.
Then, by using the supersymmetry transformation  \eqref{delta-2}, the last equation in \eqref{deltaL3-2g-2} reduces to 
\begin{align}\label{deltaL3-2g-2b}
\delta_\epsilon {\cal L}_{\rm SAH}&= m'\delta_\epsilon(\phi_1\phi_2+\bar\phi_1\bar\phi_2\big).
\end{align}
To cancel these terms, it is necessary to introduce the following inhomogeneous deformation 
\begin{align}\label{cLm}
{\cal L}_{m}=-m'(\phi_1\phi_2+\bar\phi_1\bar\phi_2\big).
\end{align}
Hence, one inhomogeneous supersymmetric Lagrangian density ${\cal L}_{{\rm SAH}m}$ is obtained by
the sum of the two Lagrangian densities ${\cal L}_{{\rm SAH}}$ \eqref{L-massive} and ${\cal L}_m$ \eqref{cLm}   \begin{align}\label{LISAH}
{\cal L}_{{\rm SAH}m}&=-\frac1{2} F_{\mu\nu}F^{\mu\nu}-i\bar\lambda\bar\sigma^\mu \partial_\mu\lambda-D_\mu\bar\phi_a D^\mu\phi_a -i\bar\psi_a\bar\sigma^\mu D_\mu\psi_a -m^2\big(|\phi_1|^2+|\phi_2|^2\big)
\nn\\
&~~~+im(\psi_1\psi_2-\bar\psi_1\bar\psi_2) -ig\Big[\lambda(\psi_1\bar\phi_1-\psi_2\bar\phi_2)-\bar\lambda(\bar\psi_1\phi_1-\bar\psi_2\phi_2)\Big] \nn\\
&~~~ -\frac14 g^2\big(|\phi_1|^2-|\phi_2|^2-\xi\big)^2 -m'\left(\phi_1\phi_2+\bar\phi_1\bar\phi_2\right).
\end{align}

 Since the global U(1) R-symmetry acts on both the supersymmetry parameter and on the dynamical fields as
\begin{align}\label{R-symmetry}
\epsilon\to e^{i\alpha}\epsilon,  \qquad\phi_a\to e^{i\alpha}\phi_a,\qquad \psi_a\to \psi_a,\qquad \lambda\to e^{i\alpha}\lambda,\qquad A_\mu\to A_\mu,
\end{align}
the projection  \eqref{Sig-epsilon} is an R-symmetry-violating constraint. Accordingly, the resulting inhomogeneous Lagrangian in \eqref{LISAH} breaks the U(1) R-symmetry which plays a  role in our discussion of the spontaneous supersymmetry breaking.

 Next, lets consider the other case with a nonzero constant mass parameter $m$, whereas  the FI coefficient $\xi$  is a function of two spatial coordinates as \begin{align} \label{mcxi}
m  \ne 0~~{\rm with}~~ \partial_{\mu} m=0, \qquad \xi=\xi(x^1,x^2).
\end{align}
Then, the variation of the Lagrangian density  \eqref{deltaL3-2g} consists only of the second term 
\begin{align}\label{deltaL3-2xi}
\delta_\epsilon {\cal L}_{\rm SAH}&=-\frac g2\Big[\partial_1\xi(\bar\epsilon\bar\sigma_{1}\lambda+\bar\lambda\bar\sigma_{1}\epsilon)+\partial_2\xi(\bar\epsilon\bar\sigma_{2}\lambda+\bar\lambda\bar\sigma_{2}\epsilon) \Big].
\end{align}
Again, in order to make the Lagrangian invariant under half supersymmetry,  halve the supersymmetry by another choice of the projection 
\begin{align}\label{ealp}
\epsilon_\alpha=\begin{pmatrix}
\chi\\
0
\end{pmatrix}\quad\Longleftrightarrow
\quad\epsilon^\alpha= \varepsilon^{\alpha \beta} \epsilon_\beta  =\begin{pmatrix}
0\\
\chi
\end{pmatrix},
\qquad\bar\epsilon_{\dot\alpha}=\begin{pmatrix}
\bar\chi\\
0
\end{pmatrix},\qquad\bar\epsilon^{\dot\alpha}= \varepsilon^{\dot \alpha \dot \beta}\bar \epsilon_{\dot \beta} =\begin{pmatrix}
0\\
\bar\chi
\end{pmatrix},
\end{align}
where $\varepsilon^{12} = \varepsilon^{\dot 1\dot 2}=1$ and $\chi$ is a complex Grassmann variable.   With this choice, the following relations are satisfied  
 \begin{align} \label{proj-2}
 \bar\epsilon\bar\sigma_2=-i\bar\epsilon\bar\sigma_1,\qquad \bar\sigma_2\epsilon=i\bar\sigma_1\epsilon.
 \end{align}
  Substitution of the projection \eqref{proj-2} in the varied Lagrangian density   \eqref{deltaL3-2xi} becomes\begin{align}\label{deltaL3-2xi2}
\delta_\epsilon {\cal L}_{\rm SAH}&=-\frac {ig}2\Big[\partial_1\xi(\bar\epsilon\bar\sigma_{2}\lambda-\bar\lambda\bar\sigma_{2}\epsilon)-\partial_2\xi(\bar\epsilon\bar\sigma_{1}\lambda-\bar\lambda\bar\sigma_{1}\epsilon) \Big]\nn\\
&=g(\partial_1\xi\delta_\epsilon A_2-\partial_2\xi\delta_\epsilon A_1)
\nonumber\\
&=g \delta_\epsilon \big[(\partial_{1}\xi) A_2 - (\partial_{2} \xi) A_1 \big],
\end{align}
where \eqref{delta-2}  have been applied in the second line. Then, the last term in \eqref{deltaL3-2xi2} can be cancelled by adding the following $B$-field deformation to the Lagrangian density  \eqref{L-massive}, 
\begin{align}\label{LabB}
{\cal L}_B=g\xi F_{12}, 
\end{align}
where $F_{12}=\partial_1 A_2-\partial_2 A_1$. The other inhomogeneous supersymmetric Lagrangian density ${\cal L}_{{\rm SAH}\xi}$ is obtained by the sum of the two Lagrangian densities ${\cal L}_{{\rm SAH}}$ \eqref{L-massive} and ${\cal L}_B$ \eqref{LabB} \begin{align}\label{SAHxi}
{\cal L}_{{\rm SAH}\xi} &=-\frac1{2} F_{\mu\nu}F^{\mu\nu}-i\bar\lambda\bar\sigma^\mu \partial_\mu\lambda-D_\mu\bar\phi_a D^\mu\phi_a-i\bar\psi_a\bar\sigma^\mu D_\mu\psi_a-m^2\big(|\phi_1|^2+|\phi_2|^2\big)
\nonumber\\
&~~~+im(\psi_1\psi_2-\bar\psi_1\bar\psi_2) -ig\Big[\lambda(\psi_1\bar\phi_1-\psi_2\bar\phi_2) -\bar\lambda(\bar\psi_1\phi_1-\bar\psi_2\phi_2)\Big]
\nonumber\\
&~~~-\frac{g^2}4 \big(|\phi_1|^2-|\phi_2|^2+g\xi F_{12}.
\end{align}
Unlike the previous projection  \eqref{Sig-epsilon}, the projection \eqref{proj-2} does not violate the R-symmetry. As a result the obtained $B$-field deformation does not  break the global U(1) R-symmetry  \eqref{R-symmetry}.
 \subsection{Superfield formalism}
 In order to figure out a simpler technique of finding the inhomogeneous extension of supersymmetric models, in this subsection, we rederive the aforementioned  inhomogeneous model in the framework of  the ${\cal N}=1$ superfield formalism.  In the superfield formalism, the Lagrangian density of the supersymmetric Abelian Higgs model with constant mass and FI coefficient consists of the K\"ahler potential $K(\Phi_a,\bar\Phi_a,V)$, the supersymmetric Abelian gauge  Lagrangian density  ${\cal L}_{\rm SAG}({\cal W})$, the holomorphic superpotential of the  complex chiral superfields $W(\Phi_i)$, and the Fayet-Iliopoulos term ${\rm {\cal L}_{FI}}(V)$, where the FI term is allowed because the gauge symmetry is Abelian. Here, $\Phi_1, \Phi_2$ are oppositely charged complex chiral superfields, $V$ is a real superfield in the Wess-Zumino gauge, and ${\cal W}_\alpha$ is the gaugino chiral superfield. Hence the Lagrangian density terms  are given by 
 \begin{align} \label{supfLg}
 & K(\Phi_a,\bar\Phi_a,V) = \bar\Phi_1 e^{-2gV}\Phi_1 +\bar\Phi_2 e^{2gV}\Phi_2, \nn\\
 &{\cal L}_{\rm SAG}({\cal W}) + {\rm c.c.} = -\frac1{2} ~{\cal W}^\alpha{\cal W}_\alpha -\frac1{2} ~\bar{\cal W}^{\dot\alpha}\bar{\cal W}_{\dot\alpha},\nn\\
 &W(\Phi_a)+{\rm c.c.} = -im \Phi_1\Phi_2 +im \bar\Phi_1\bar\Phi_2,\nn\\
 &  {\cal L}_{\rm FI}(V)= 2g\xi V.
 \end{align}
 The superspace expansions of superfields are
 \begin{align}\label{supExp}
 &\Phi_a(y,\theta)=\phi_a(y)+\sqrt2\theta\psi_a(y)+\theta^2 f_a(y^\mu),~\nn\\
 &V(x,\theta,\bar\theta)=\theta\sigma^\mu\bar\theta A_\mu(x)+\frac1{\sqrt2}\left[ i\theta^2\bar\theta\bar\lambda(x)-i\bar\theta^2\theta\lambda(x)\right]+\frac12\theta^2\bar\theta^2 d(x),\nn\\
 &{\cal W}_\alpha(y,\theta)=\frac1{\sqrt2}\lambda_{\alpha}(y)-\sigma^{\mu\nu~\beta}_\alpha\theta_\beta F_{\mu\nu}(y)+i\theta_\alpha d(y)+\frac{i}{\sqrt2}\theta^2\sigma^{\mu}_{\alpha\dot\alpha}\partial_\mu\bar\lambda^{\dot\alpha}(y),
\end{align}  
where $y^\mu=x^\mu+i\theta\sigma^\mu\bar\theta$ with Grassmann odd coordinates $\theta$, $\bar\theta$, and  $f_a$,  $d$ are auxiliary fields.  The corresponding off-shell supersymmetric variations of these component fields  are
\begin{align} \label{susysupf}
&\delta A_\mu=-\frac{\sqrt2}{2}(i\bar\epsilon\bar\sigma_\mu\lambda-i\bar\lambda\bar\sigma_\mu\epsilon),\quad\delta\lambda=\sqrt2(-\sigma^{\mu\nu}\epsilon F_{\mu\nu}+i\epsilon d),\quad \delta d=\frac{\sqrt2}2(\bar\epsilon\bar\sigma^\mu\partial_\mu\lambda-\epsilon\sigma^\mu\partial_\mu\bar\lambda)\nn\\
&\delta\phi_a=\sqrt2\epsilon\psi_a, \quad \delta\psi_a =i\sqrt2\sigma^\mu\bar\epsilon D_\mu\phi_a+\sqrt2\epsilon f_a,\quad \delta f_a=i\sqrt2\bar\epsilon\bar\sigma^\mu D_\mu\psi_a. 
\end{align}

The integration of \eqref{supfLg} over the Grassmann odd coordinates $(\theta,\bar\theta)$ results in a  Lagrangian density which is the sum of the last $f$ or $d$ term in the superspace expansions of the various composite superfields. The K\"ahler and the FI terms in \eqref{supfLg} are real composite/single superfields, and thus their last terms are $d$-terms. On the other hand, the  Lagrangian density of supersymmetric Abelian gauge theory and the holomorphic superpotentials are chiral composite superfields, and thus their last terms are $f$-terms. From \eqref{susysupf}, we note that the variations of the $d$-terms and  $f$-terms of the superfields are expressed by total derivatives
\begin{align}\label{dfdd}
\delta f=i\bar\epsilon\bar\sigma^\mu\partial_\mu C_{\theta}, \qquad \delta d=i\bar\epsilon\bar\sigma^\mu\partial_\mu C_{\bar\theta^2\theta}+i\epsilon\sigma^\mu\partial_\mu C_{\theta^2\bar\theta},
\end{align}     
and hence the action built from the terms in \eqref{supfLg} is invariant. Here, $C_{\theta}$ is the coefficient of the $\theta$ term in the superspace expansion of a composite chiral superfield, whereas $C_{\bar\theta^2\theta}$ and $C_{\theta^2\bar\theta}$  are the coefficients of the ${\bar\theta^2\theta}$ and ${\theta^2\bar\theta}$ terms, respectively, in the superspace expansion of a composite real superfield. Since the coefficients  $C_\theta$, $C_{\bar\theta^2\theta}$, and $C_{\theta^2\bar\theta}$, are invariant under the gauge transformation because $f$ and  $d$ are the last terms of some gauge-invariant composite chiral or real superfields,  ordinary derivatives are used  instead of covariant derivatives in \eqref{dfdd}.

Consider the first case of inhomogeneity by turning  on the inhomogeneity in the model of our consideration through the mass parameter $m$  \eqref{supfLg} with dependence on spatial coordinates.  With the inhomogeneous mass, the fact that the variation of the $f$-term is expressed by total derivatives does not mean that the variation of the superpotential is a total derivative. In order to exhibit this clearly,  expand the composite chiral  superfield $\Phi_1 \Phi_2$ as     
\begin{align}
\Phi_1 \Phi_2=\phi_1\phi_2+\sqrt2\theta(\psi_1\phi_2+\phi_1\psi_2)+\theta^2(f_1\phi_2+\phi_1 f_2-\psi_1\psi_2).
\end{align}
Then,  the coefficient $C_\theta$ for this composite chiral superfield is $\sqrt2(\psi_1\phi_2+\phi_1\psi_2)$, so that the variation of the superpotential involves non-total derivative term because of the inhomogeneity of the mass $\partial_{\mu} m \ne 0$
\begin{align}
\int d^2\theta~\delta W(\Phi_a)+ {\rm c.c.} &=m\sqrt2  \bar\epsilon\bar\sigma^\mu\partial_\mu (\psi_1\phi_2+\phi_1\psi_2)+{\rm c.c.}\nn\\
&=-(\partial_\mu m)\sqrt2  \bar\epsilon\bar\sigma^\mu(\psi_1\phi_2+\phi_1\psi_2)+{\rm c.c.}+{\rm total ~derivative}.
\end{align} 
If we assume for physics purpose that the mass function depends only on a single coordinate $x^1$ and the projection  \eqref{Sig-epsilon} is employed, then we obtain
\begin{align}\label{dW}
\int d^2\theta~\delta W(\Phi_a)+c.c.&= m'\sqrt2  (\epsilon\psi_1\phi_2+\phi_1\epsilon\psi_2)+ {\rm c.c.}+{\rm total ~derivative}\nn\\
&= m'\delta(\phi_1\phi_2)+ {\rm c.c.}+{\rm total ~derivative}.
\end{align} 
The first non-trivial derivative term means that the supersymmetric invariance is violated by the assumed inhomogeneous mass function $m= m(x^1)$.  
To cancel the offending term in \eqref{dW}, it is required to add the inhomogeneous mass deformation to the Lagrangian \begin{align}
{\cal L}_{m}=-m' (\phi_1\phi_2 + \bar \phi_1 \bar{\phi}_2)
\end{align}
which is written in superfield formalism as
\begin{align}
{\cal L}_m =-m' \Big(\int d^2\theta~\theta^2\Phi_1\Phi_2 + \int d^2\bar\theta~\bar\theta^2\bar\Phi_1 \bar\Phi_2\Big).
\end{align}
Now the first inhomogeneous supersymmetric Lagrangian density ${\cal L}_{{\rm SAH}m}$ of ${\cal N} = 1$ \eqref{LISAH} is reproduced in superfield formalism. 

Next, consider the second case of inhomogeneity by taking into account  the inhomogeneous FI coefficient. In the superspace expansions of superfields  \eqref{supExp}, the coefficients $C_{\bar\theta^2\theta}$ and $C_{\theta^2\bar\theta}$ are read from the superspace expansion of the real superfield $V$ and one can write \begin{align}
\int d^2\theta d^2\bar\theta~\delta {\rm {\cal L}_{FI}}(V)=-\frac{g}{\sqrt2}\partial_\mu\xi\big(\bar\epsilon\bar\sigma^\mu\lambda-\epsilon\sigma^\mu\bar\lambda\big)+{\rm total ~ derivative}.
\end{align}
If we regard the coefficient  $\xi$ in the FI term as a function of  two coordinates $(x^1, x^2)$ and impose the projection   \eqref{proj-2}, then we obtain
again non-trivial derivative terms \begin{align}\label{dLFI}
\int d^2\theta d^2\bar\theta~\delta {\rm {\cal L}_{FI}}(V)&=-\frac{g}{\sqrt2}\partial_1\xi\big(i\bar\epsilon\bar\sigma^2\lambda-i\bar\lambda\bar\sigma^2\epsilon\big)+\frac{g}{\sqrt2}\partial_2\xi\big(i\bar\epsilon\bar\sigma^1\lambda-i\bar\lambda\bar\sigma^1\epsilon\big)+{\rm total ~ derivative}\nn\\
&= g(\partial_1\xi\delta A_2-\partial_2\xi\delta A_1)+{\rm total ~ derivative}.
\end{align}
The supersymmetry-breaking  terms in  \eqref{dLFI}  can be removed by adding
the following terms to the Lagrangian  \begin{align}
{\cal L}_{\xi}=-g(\partial_1\xi A_2-\partial_2\xi A_1)
\end{align}
which is expressed in superfield formalism  as
\begin{align}
{\cal L}_{\xi} =2g\int d^2\theta d^2\bar\theta \big[ (\partial_1\xi)~\theta \sigma_2\bar\theta~ V- (\partial_2\xi)~ \theta \sigma_1\bar\theta~ V\big].
\end{align}
 Hence, the second inhomogeneous supersymmetric Lagrangian density ${\cal L}_{{\rm SHA}\xi}$ of ${\cal N}=1$ \eqref{SAHxi} is also rederived in superfield formalism.  
 
 In general, the ${\cal N}= 1$ superfield formalism is used to write the Lagrangian density of any homogeneous supersymmetric field theory  as ${\cal L}\sim \sum_r  C_{\rm last}^r $ in which  $C_{ \rm last}^r$'s are the coefficients of the highest order terms in the $\theta$, $\bar \theta$ expansions for some composite/single superfields. Then, the supersymmetric variation can be written in terms of parameters of the theory $\lambda_r$
and the  coefficients of the terms preceding the highest order terms in the $\theta$, $\bar \theta$ expansions $C_{\rm last -1}^{r}$ 
\begin{align}\label{dLC}
\delta \mathcal{L} \sim \sum_r \lambda_r \partial_{\mu} C_{\rm last -1}^{r}.
\end{align}
This variation is  a total derivative if the parameters of the theory are constants and thus the theory is  invariant under supersymmetric transformations. If the parameters of the theory depend on spatial coordinates,  then integration by parts of \eqref{dLC} and assignment of an appropriate projection on the supersymmetry parameter will produce  inhomogeneous deformations that are required to make the theory  partially supersymmetric.

\section{Supersymmetric Solutions}

In order to motivate the discussion of spontaneous supersymmetry breaking, in this section we establish the BPS limits for the inhomogeneous field theory models of the previous section and find some static  solutions to the corresponding BPS equations.

\subsection{Killing spinor equations}
Once a supersymmetric field theory is constructed, it is connected to the BPS structure, i.e. BPS limit, BPS equations, and BPS objects. In relation with the BPS equations,  it is  systematic to derive them from the Killing spinor equations. 

 First  the BPS equations are derived for the case of inhomogeneous mass parameter, directly from the supersymmetry transformations, by
requiring vanishing supersymmetry variations of the fields. 
Lets apply the  projection \eqref{Sig-epsilon} and  write the variation of the fermionic fields $\psi_i$ as 
\begin{align}\label{delta-f1}
\delta_\epsilon\psi_a &=i\sigma^\mu\bar\epsilon D_\mu\phi_a-im_{ab}\sigma^1\bar\epsilon \bar\phi_b
\nonumber\\
&
=\Big[\sigma^1\big(i D_1\phi_a-im_{ab} \bar\phi_b\big)+i\sigma^n D_n\phi_a\Big]\bar\epsilon, \qquad (n={0,2,3}).
\end{align}
In order for the Killing spinor equation $\delta_\epsilon\psi_a=0$, 
the square parenthesis in the above expression \eqref{delta-f1} must vanish. Since the $\sigma^\mu$ matrices are linearly independent, the following two equations must hold
independently 
\begin{align}\label{BPS1}
&D_1\phi_a -m_{ab}\bar\phi_b =0,\qquad D_n\phi_a=0.
\end{align}

Next, the variation of the gaugino $\lambda$ is expressed as 
\begin{align}
\delta_\epsilon\lambda_\alpha& =- (\sigma^{\mu\nu})_\alpha^{~\beta}\epsilon_\beta F_{\mu\nu}+\frac{ig}2\big(|\phi_1|^2-|\phi_2|^2 -\xi\big)\epsilon_\alpha.
\end{align}
Inserting $\alpha=1$ and $\alpha=2$, we obtain
\begin{align}\label{deltaL1-2}
&\delta_\epsilon\lambda_1= \Big[F_{03}+iF_{12}+\frac{ig}2\big(|\phi_1|^2-|\phi_2|^2-\xi\big)\Big]\epsilon_1+ \Big[F_{01}-iF_{02}-F_{13}+iF_{23}\Big]\epsilon_2,\nn\\
&\delta_\epsilon\lambda_2 = \Big[F_{01}+iF_{02}+F_{13}+iF_{23}\Big]\epsilon_1 + \Big[-F_{03}-iF_{12}+\frac{ig}2\big(|\phi_1|^2-|\phi_2|^2-\xi\big)\Big]\epsilon_2.
\end{align}
Both the real and imaginary parts of all the square parentheses  in the equations \eqref{deltaL1-2} should vanish in order for the Killing spinor equation $\delta_\epsilon\lambda_\alpha=0$. Then, we obtain
\begin{align}\label{BPS2}
&\ F_{\mu\nu}=0,\qquad |\phi_1|^2-|\phi_2|^2-\xi=0.
\end{align} 
Hence, for the supersymmetric Abelian Higgs model with the inhomogeneous mass parameter $m = m(x^1)$ \eqref{mux} of the Lagrangian ${\cal L}_{{\rm SAH}m}$ \eqref{LISAH}, the Killing spinor equations lead to the two sets of the BPS equations \eqref{BPS1} and \eqref{BPS2}. 

Similarly, for the case of inhomogeneous FI coefficient,  the projection $\sigma^3\bar\epsilon=-\sigma^0\bar\epsilon,~\sigma^2\bar\epsilon=-i\sigma^1\bar\epsilon$ in \eqref{proj-2} is applied in order to simplify the variation of the chiral fermion $\delta_\epsilon\psi_i$ as follows
\begin{align}
\delta_\epsilon\psi_a &=i\Big[\sigma^0\big( D_0\phi_a-D_3\phi_a\big)+\sigma^1\big( D_1\phi_a-iD_2\phi_a\big)\Big]\bar\epsilon-i m_{ab}\bar\phi_b \epsilon.
\end{align}
Requirement of a Killing spinor equation $\delta_\epsilon\psi_a =0$ with independence of two supersymmetry parameters $\epsilon$ and $\bar\epsilon$  leads to the following BPS equations, 
\begin{align}\label{BPSxi1}
 (D_0 -D_3 )\phi_a=0,\qquad (D_1 -iD_2)\phi_a =0,\qquad m_{ab} \bar\phi_b=0.
\end{align}
The gaugino variation $\delta_\epsilon\lambda$    \eqref{deltaL1-2} is simplified by substitution of  $\epsilon_1=\chi$ and $\epsilon_2=0$ in \eqref{ealp} as
\begin{align}\label{deltaLxi}
&\delta_\epsilon\lambda_1=  \Big[F_{03}+iF_{12}+\frac{ig}2\big(|\phi_1|^2-|\phi_2|^2-\xi\big)\Big]\chi,\nn\\
&\delta_\epsilon\lambda_2= \Big[F_{01}+iF_{02}+F_{13}+iF_{23}\Big]\chi.
\end{align}
 Both the real and imaginary parts of each square parenthesis in \eqref{deltaLxi} should vanish in order for the Killing spinor equation $\delta_\epsilon\lambda_\alpha=0$. Then, we obtain the following BPS equations,
\begin{align}\label{BPSxi2}
F_{03}=0,\qquad F_{12}+\frac{g}2\big(|\phi_1|^2-|\phi_2|^2-\xi\big)=0,\qquad F_{01}+F_{13}=0,\qquad F_{02}+F_{23}=0.
\end{align}
For the other supersymmetric Abelian Higgs model with the inhomogeneous FI coefficient $\xi = \xi(x^1, x^2)$ \eqref{mcxi} of the Lagrangian ${\cal L}_{{\rm SAH}\xi}$ \eqref{SAHxi}, the Killing spinor equations lead to the other two sets of the BPS equations \eqref{BPSxi1} and \eqref{BPSxi2}.  

\subsection{ BPS limit}
Another method to derive the BPS equations is to consider the  Hamiltonian of the bosonic part, to rewrite it as the sum of complete squares and a boundary term by reshuffling the terms, and to saturate the limit by setting each of the complete squares equal
to zero. In this subsection, bosonic sectors of the two inhomogeneous supersymmetric Abelian Higgs models of the Lagrangian densities ${\cal L}_{{\rm SAH}m}$ \eqref{LISAH} and ${\cal L}_{{\rm SAH}\xi}$ \eqref{SAHxi} are taken into account to rederive the Bogomolny equations.

To apply this procedure first to the inhomogeneous Abelian Higgs model with inhomogeneous mass function,  turn off the fermionic fields to zero $\lambda = \psi_a = 0$ in the Lagrangian density ${\cal L}_{{\rm SAH}m}$  \eqref{LISAH} and compute the bosonic  energy-momentum tensor\begin{align}
T_{\mu\nu}
&=2D_\mu\bar\phi_a D_\nu\phi_a+ 2 F_{\mu\rho}F_\nu^{~\rho}+g_{\mu\nu}\Big[-D_\rho\bar\phi_a D^\rho\phi_a-\frac1{2 } F_{\rho\rho'}F^{\rho\rho'}\nn\\
&~~~ -\frac14 g^2\big(|\phi_1|^2-|\phi_2|^2-\xi\big)^2-m^2\big(|\phi_1|^2+|\phi_2|^2\big)-m'\big(\phi_1\phi_2+\bar\phi_1\bar\phi_2\big)\Big].\end{align}
After some rearrangement reflecting the BPS equations \eqref{BPS1} and \eqref{BPS2}, the Hamiltonian density obtained from the energy density ${\cal H}=T_{00}$ can be written by the sum of absolute square terms and a boundary term $K'=\frac{dK}{dx}$
\begin{align}\label{energy}
{\cal H}&= \frac1{2}|F_{\mu\nu}|^2 +\big|D_1\phi_1-m\bar\phi_2\big|^2+\big|D_1\phi_2-m\bar\phi_1\big|^2\nn\\
&~~ + \sum_{n=0,2,3}\Big(\big|D_n\phi_1\big|^2+\big|D_n\phi_2\big|^2\Big)+\left[\frac{g}{2}\big(|\phi_1|^2-|\phi_2|^2-\xi\big)\right]^2+ K',
\end{align}
where  $K=m(\bar\phi_1\bar\phi_2+\phi_1\phi_2)$.
When the Hamiltonian density \eqref{energy}  is bounded by the spatial total derivative term ${\cal H}\ge K'$, the BPS equations  in \eqref{BPS1} and \eqref{BPS2}, which can be read from the absolute square terms,
 are satisfied.
Since $K'$ is not positive semidefinite and the corresponding energy $E = \int d^3x {\cal H}$ can possibly have arbitrary negative value including negative infinity, the inequality ${\cal H}\ge K'$ from
\eqref{energy}  does not guarantee existence of the minimum value of the Hamiltonian.

Since the field strength tensor  $F_{\mu\nu}$ vanishes in \eqref{BPS2}, the gauge field $A_\mu$ becomes a pure gauge degree of freedom. Accordingly, the covariant derivatives $D_\mu$ can be replaced by ordinary derivatives $\partial_\mu$, and the remaining BPS equations consist only of the first order scalar equations and constraints 
\begin{align}\label{BPS-2}
&\partial_1 \phi_1-m\bar\phi_2=0,\qquad \partial_1 \phi_2-m\bar\phi_1=0,\qquad \partial_n\phi_1=0,\nn\\
&\partial_n\phi_2=0,\qquad|\phi_1|^2-|\phi_2|^2-\xi=0,\qquad F_{\mu\nu}=0, \qquad  (n = 0,2,3).
\end{align}

In order to check the consistency of the BPS equations in \eqref{BPS1} and \eqref{BPS2} with the equations of motion of the scalar fields, we need to apply one more covariant derivative to each of the BPS equations in \eqref{BPS1}  to obtain the following equations \begin{align}\label{BPS-3}
&D^1 D_1\phi_1-m'\bar\phi_2-m^2\phi_1=0, \qquad D^1D_1\phi_2-m'\bar\phi_1-m^2\phi_2=0, \nonumber\\
& D^nD_n\phi_1=0, \qquad D^nD_n\phi_2=0.
\end{align}
After turning off  the fermionic fields in the Lagrangian density ${\cal L}_{{\rm SAH}m}$ \eqref{LISAH},    the Euler-Lagrange equations  for the scalar fields are reproduced from \eqref{BPS-3} with the help of the constraint equation of the scalar fields in \eqref{BPS2} as
\begin{align}\label{EOM38}
&\big(D^1 D_1\phi_1-m'\bar\phi_2-m^2\phi_1\big)+D^n D_n\phi_1-\frac{g^2}{2}(|\phi_1|^2-|\phi_2|^2-\xi)\phi_1=0, \nn\\
&\big(D^1 D_1\phi_2-m'\bar\phi_1-m^2\phi_2\big)+D^n D_n\phi_2+\frac{g^2}{2}(|\phi_1|^2-|\phi_2|^2-\xi)\phi_2=0.
\end{align}
In addition,  to show consistency with the Euler-Lagrange equation for the gauge field  
\begin{align}\label{ELgauge}
 2 \partial^\mu F_{\mu\nu}+i\Big(D_\nu\bar\phi_1\phi_1-\bar\phi_1D_\nu\phi_1-D_\nu\bar\phi_2\phi_2+\bar\phi_2D_\nu\phi_2\Big)=0,
\end{align} 
substitution of the BPS equations  in \eqref{BPS2} with the condition of static fields is enough. 
Therefore,  any BPS solution of the equations in 
\eqref{BPS1} and \eqref{BPS2}
is a solution of the Euler-Lagrange equations \eqref{EOM38} as well. However, it does not mean that an arbitrary static pure scalar solution of the Euler-Lagrange equations \eqref{EOM38} and \eqref{ELgauge} obeys the BPS equations in \eqref{energy} and \eqref{BPS-2},
and thus possible existence of non-BPS solution is an open question.

Now it is the turn of the second inhomogeneous supersymmetric Abelian Higgs model with the inhomogeneous FI coefficient whose dynamics is governed by the Lagrangian density ${\cal L}_{{\rm SAH}\xi}$ \eqref{SAHxi}. Set all the fermions to be zero $\lambda = \psi_a = 0$, read the bosonic Lagrangian density, and find the bosonic energy-momentum density tensor 
\begin{align}
T_{\mu\nu}&=2D_\mu\bar\phi_a D_\nu\phi_a +  F_{\mu\rho}F_\nu^{~\rho}+g_{\mu\nu}\Big[-D_\rho \bar\phi_a D^\rho\phi_a-\frac1{2} F_{\rho\rho'}F^{\rho\rho'}-\frac14 g^2 \big(|\phi_1|^2-|\phi_2|^2-\xi\big)^2\nn\\
&~~~-m^2\big(|\phi_1|^2+|\phi_2|^2\big)+g \xi F_{12}\Big].\end{align}
From this energy-momentum tensor, obtain the Hamiltonian density ${\cal H}$  as the energy density ${\cal H} = T_{00}$ and make the sum of absolute square terms except for the total divergence term after reshuffling the terms with the help of integrations by parts and $[D_1, D_2] \phi_a = -(-1)^a ig F_{12} \phi_a$
\begin{align}
{\cal H} &= F_{i0}^2 + \frac{1}{2} F_{ij} + |D_0 \phi_a|^2 + |D_i \phi_a|^2 + m\big( |\phi_1|^2 + \phi_2|^2 \big)
\nonumber\\
&~~~+ \frac{g^2}{4} \big(|\phi_1|^2 - |\phi_2|^2 - \xi({\bf x}) \big)^2 - g \xi F_{12} \nonumber\\
&= (F_{10} - F_{13})^2 + (F_{20} - F_{23})^2 + F_{03}^2 + \big|(D_0 - D_3) \phi_a \big|^2 + \big|(D_1 - i D_2) \phi_a \big|^2 \nonumber\\
&~~~ +\big[ F_{12} + \frac{g}{2} \big( |\phi_1|^2 - |\phi_2|^2 - \xi({\bf x})\big) \big]^2 + |m \phi_1 |^2 + |m \phi_2|^2
\nonumber\\
&~~~+ 2 F_{10} F_{13} + 2 F_{20} F_{23} + D_0 \bar \phi_a D_3 \phi_a + D_3 \bar{\phi_a} D_0 \phi_a - \frac{1}{2} \epsilon^{\bar i \bar j} \partial_{\bar{i}} j^{\bar{j}} ],
\end{align}
where ${\bf x} = (x,y)$, the indices $\bar{i}$, $\bar{j}$ run over 1 and 2,  and the U(1) current $j^\mu$ is defined by
\begin{align}
 j^\mu =-i \sum_{a=1}^2 \big( \bar{\phi}_a D^\mu \phi_a - D^\mu \bar{\phi}_a \phi_a \big).
\end{align}
Once all the BPS equations in \eqref{BPSxi1} and \eqref{BPSxi2}  are satisfied, all the absolute square terms vanish and thus an equality holds
\begin{align}
{\cal H} = 2 F_{10}^2 + 2 F_{20}^2 + 2 |D_0 \phi_a|^2 - \frac{1}{2} \epsilon^{\bar i \bar j} \partial_{\bar{i}} j^{\bar{j}}.
\end{align}
For static BPS objects, the Weyl gauge $A^0 = 0$ is chosen and applied. Hence electric field of such BPS objects vanish $F_{i0} = 0$ with time component of covariant derivative $D_0 \phi_a = 0$. In summary, any static BPS objects of an inhomogeneous supersymmetric Abelian Higgs model of the Lagrangian density ${\cal L}_{{\rm SAH} \xi}$ \eqref{SAHxi} saturates the following BPS limit in terms of its energy of bosonic sector 
\begin{align}\label{xiE}
E & = \int d^4 x \Big\{ |F_{13}|^2 + |F_{23}|^2 + |D_3 \phi_a |^2 + |m \phi_1|^2 + |m \phi_2|^2 \nonumber\\
&~~~~~~~~~~~~~~ +\big|(D_1 - i D_2) \phi_a\big|^2 + \big[ F_{12} + \frac{g}{2} \big(|\phi_1|^2 - |\phi_2|^2 - \xi({\bf x})\big) \big]^2 \Big\}
\nonumber\\
&~~~~~~~~~~~~~~+ \frac{1}{2} \int_{-\infty}^{\infty } dz \oint_{\partial {\mathbb R}^2_{xy}} dx^{\bar{i}} j^{\bar{i}},    
\end{align}
where $\partial {\mathbb R}^2_{xy}$ denotes the spatial boundary $\sqrt{x^2 + y^2} \to \infty $ of the $xy$-plane. In the first line the five absolute square terms become zero by the trivial solutions of the Bogomolny equations, and, in the second line, the two absolute square terms lead to the two remaining nontrivial Bogomolny equations 
\begin{align}
&(D_1 - i D_2) \phi_a = 0,
\nonumber\\
&F_{12} = - \frac{g}{2} \big( |\phi_1|^2 - |\phi_2|^2 - \xi({\bf x}) \big).
\end{align}
In the third line, $x$- and $y$-components of the U(1) current decrease rapidly to zeor at infinity $\lim_{\sqrt{x^2 + y^2} \to \infty} j^{\bar i} = 0$ and thus the boundary integral is zero.

In order to check the consistency between the BPS equations in \eqref{BPSxi1} and \eqref{BPSxi2} and the  equations of motion, apply one more covariant derivative to the BPS equations and  rearrange the equations \begin{align}\label{DmuBPS}
&D^0\big(D_0\phi_a -D_3\phi_a \big)-D^3\big(D_0\phi_a -D_3\phi_a\big)=0 \Longrightarrow D^0D_0\phi_a +D^3D_3\phi_a =0,\nn\\
&D^1\big(D_1\phi_a-iD_2\phi_a\big)+iD^2\big(D_1\phi_a-iD_2\phi_a\big)=0 \Longrightarrow D^1D_1\phi_a +D^2D_2\phi_a =0,\nn\\
&\bar\phi_a =0.
\end{align}
On the other hand, with the homogeneous mass, the Euler-Lagrange equations for the scalar fields is 
\begin{align}
D^\mu D_\mu\phi_a -m^2\phi_a\pm\frac{g^2}{2}(|\phi_1|^2-|\phi_2|^2-\xi)\phi_a =0
\end{align}
which are proven to be consistent with the rearranged BPS equations in \eqref{DmuBPS}. 
For the gauge field, the Euler-Lagrange equation in \eqref{ELgauge} is modified by the last term in the Lagrangian density ${\cal L}_{{\rm SAH}\xi}$ \eqref{SAHxi}
\begin{align}
2\partial^\nu F_{\nu\mu}+i\Big( D_\mu\bar\phi_1\phi_1-\bar\phi_1D_\mu \phi_1-D_\mu\bar\phi_2 \phi_2+\bar\phi_2D_\mu \phi_2\Big)+g\partial^\nu\xi\big(\delta_{1\nu} \delta_{2\mu} -\delta_{2\nu} \delta_{1\mu}\big)=0
\end{align}
which is also satisfied by the BPS equations if we set $F_{\mu\nu}=-\frac{g}{2}\big(\delta_{\mu 1}\delta_{\nu 2}-\delta_{\nu1}\delta_{\mu2}\big) \xi(x^1, x^2)$.

\subsection{BPS vacuum solutions}
Now we have the two sets of the Bogomolny equations including two species of inhomogeneity: One is the inhomogeneous mass function $m= m(x^1)$ and the other is the inhomogeneous FI coefficient $\xi = \xi(x^1, x^2)$. These inhomogeneities possibly do not allow
homogeneous BPS vacuum solutions, rather there may support inhomogeneous BPS solutions of zero energy. Hence, in the current subsection, we examine the Bogomolny equations and find the nontrivial solution of zero energy as new inhomogeneous BPS vacuum configurations. 

Consider the first case in which  the inhomogeneous mass function $m=m(x^1)$ is turned on through the inhomogeneous mass deformation \eqref{cLm}.  In order to find a vacuum solution, an ansatz compatible to the Bogomolny equations in \eqref{BPS1} and \eqref{BPS2} is taken, that the two scalar fields  have opposite phases
\begin{align} \label{h12}
\phi_1= h_1 e^{i\alpha},\qquad \phi_2= h_2e^{-i\alpha}.
\end{align}
Substitution of the ansatz \eqref{h12} into the Bogomolny equations in   \eqref{BPS1} and \eqref{BPS2} forces that the phase  $\alpha$ should be  a constant. Hence the BPS equations in \eqref{BPS1} and \eqref{BPS2}  with inhomogeneous mass are simplified as the equations of scalar amplitudes $h_1$ and $h_2$ with zero electromagnetic field\begin{align} \label{BPS-az}
h_1'-m h_2=0,\qquad h_2'-m h_1=0,\qquad h_1^2-h_2^2-\xi=0,\qquad F_{\mu\nu}=0.
\end{align}
Using the boundary function   
\begin{align}\label{bndyK}
K=2mh_1h_2
\end{align}
we can rewrite the first two equations in \eqref{BPS-az} in terms of $\frac{\partial K}{\partial h_a}$, 
\begin{align}
h'_1-\frac12\frac{\partial K}{\partial h_1}=0,\qquad h'_2-\frac12\frac{\partial K}{\partial h_2}=0,\qquad h_1^2-h_2^2-\xi=0,\qquad F_{\mu\nu}=0.
\end{align}
The homogeneous solution $h_a=0$ to the first two BPS equations is just an extremum of the boundary function $K$, however the trivial solution does not satisfy the third BPS equation for nonzero parameter $\xi\ne0$. 

 For the model with inhomogeneous mass  function \eqref{cLm} and homogeneous FI coefficient, the R-symmetry is broken and we expect to obtain supersymmetric solutions to the BPS equations in \eqref{BPS-az}. Indeed, the inhomogeneous supersymmetric solutions are obtained in explicit form 
\begin{align}\label{soln1}
&h_1=\frac12\Big[\big(c+\sqrt{c^2-\xi}\big)e^{a(x^1)}+\frac{\xi}{c+\sqrt{c^2-\xi}} e^{-a(x^1)}\Big],\nn\\ 
&h_2=\frac12\Big[\big(c+\sqrt{c^2-\xi}\big)e^{a(x^1)}-\frac{\xi}{c+\sqrt{c^2-\xi}} e^{-a(x^1)}\Big],
\end{align}
where $a(x^1)=\int_{x_0}^{x^1} m(x)dx ~~ {\rm and}~~c=h_1(x_0)$.
Note that $c$ is a positive integration constant and $c^2\ge\xi$. 
Then, the boundary function $K$ \eqref{bndyK} is computed to be  
\begin{align}
&K=\frac m2 \Big[\big(c+\sqrt{c^2-\xi}\big)^2e^{2a(x^1)}-\frac{\xi^2}{\big(c+\sqrt{c^2-\xi}\big)^2} e^{-2a(x^1)}\Big].
\end{align}
In order to find the solutions that are  finite  in asymptotic regions $x^1\to\pm \infty$, the mass function $m(x^1)$ in \eqref{cLm} should  vanish in the asymptote  that leads to the zero boundary function $K(\infty)=K(-\infty)=0$. Accordingly the vacuum energy $E_{\rm v}$ is zero as well
\begin{align}\label{Ev}
E_{\rm v} =K(\infty)-K(-\infty)=0.
\end{align}
Note that these solutions are not valid in the case of constant mass function, because in that case, the model is a spontaneous supersymmetry breaking model and supersymmetric vacuum solutions can not be found. Furthermore, mass functions which asymptotically approach finite constants result in diverging solutions with divergent energy and are not suitable.  
Explicit form of the exact inhomogeneous BPS vacuum  solutions of zero energy  \eqref{Ev} can be written by choosing a specific mass function $m(x^1)$ that vanishes at asymptotic boundaries. Examples of such mass function include Gaussian  function, Dirac delta  function, etc.

 In the case of inhomogeneous FI coefficient $\xi(x^1, x^2)$ in \eqref{mcxi}, an almost trivial BPS solution is readily obtained as
\begin{align}\label{xiBPS}
\phi_a =0,\qquad F_{\mu\nu}=-\frac{g}{2} \big(\delta_{\mu 1}\delta_{\nu 2}-\delta_{\nu1}\delta_{\mu2}\big)\xi(x^1, x^2).
\end{align}
Inserting the solution into the energy \eqref{xiE}, all the absolute square terms in the first and second lines vanish and zero spatial current components $j^{\bar i} =0$ removes the term in the third line. Therefore, the energy \eqref{xiE} for the BPS configuration \eqref{xiBPS} is bounded as 
\begin{align}
E \ge  0.
\end{align}
 We now confirm that the obtained solution
\eqref{xiBPS} is a inhomogeneous vacuum configuration of zero energy. We note that, though the R-symmetry is unbroken, the existence of the exact nontrivial vacuum solutions means the supersymmetry is not spontaneously broken in the inhomogeneous model with the inhomogeneous FI coefficient \eqref{mcxi}. We will discuss more about this point in the subsequent section.

The reason why the nontrivial inhomogeneous BPS vacuum solution \eqref{mcxi} of zero energy exists but the homogeneous vacuum configuration can not be sustained as a BPS solution in the presence of the homogeneous FI term is spontaneous breakdown of the supersymmetry. This consequence is actually expected, since the homogeneous model is R-symmetry invariant supersymmetric model with spontaneously broken R-symmetry, and hence it is a spontaneous supersymmetry breaking model by the Nelson-Seiberg argument~\cite{Nelson:1993nf}. This is actually the  D-term breaking model\cite{Fayet:1974jb} and we will discuss more about it in the next section. 
The existence of inhomogeneous supersymmetric vacuum solutions of zero energy means the inhomogeneous model is no more  a spontaneous supersymmetry breaking model.

\section{Spontaneous Supersymmetry Breaking in Inhomogeneous Models }\label{sec-4}
 For convenience, let us  briefly recapitulate  the Nelson-Seiberg argument~\cite{Nelson:1993nf} on the spontaneous supersymmetry breaking. For ${\cal N}=1$ supersymmetric theory of $n$ chiral superfields $\Phi_a$, the supersymmetry is spontaneously broken if there is no solution to the $f$-term equation \begin{align} \label{ftermeq}
 \frac{\partial W(\Phi_a)}{\partial\Phi_a}=0, \quad {\rm for} ~a=1,\dots, n,
 \end{align}
 where $W(\Phi_a)$ is the holomorphic superpotential.  For a generic superpotential, the relations in \eqref{ftermeq} are $n$ equations for $n$ independent variables $\Phi_a$, and it seems there always exists a supersymmetric vacuum solution.\footnote{ The superpotential is referred to as `generic' if it is a polynomial function of degree $k$ and incorporates all terms of degree $k$ or lower, that are not prohibited by the symmetries of the theory.} However, the situation changes, if the theory is invariant under the global U(1) R-symmetry transformation. Recalling that the superpotential is charged under the R-symmetry with R-charge $R[W] = 2$, it can be expressed in terms of  R-symmetry invariant ratios $\big(X_a=\frac{\Phi_a}{\Phi_1^{r_a/r_1}},~~a=2,\dots n\big)$ as 
 \begin{align}
 W=\Phi_1^{2/r_1} W(X_a),
 \end{align}
 where $r_a$ is the R-charges of chiral superfields $\Phi_a$. Then, the $n$ $f$-term equations  \eqref{ftermeq} become 
\[ \frac{\partial W}{\partial\Phi_a}= \left\{ \begin{array}{ll}
         \Phi_1^{(2-r_a)/r_1}\frac{\partial W}{\partial X_a}=0,& \mbox{if $a \ne 1$},\\
         \frac{2}{r_1}\Phi_1^{(2-r_1)/r_1}W+\Phi_1^{2/r_1}\frac{\partial W}{\partial X_b}\frac{\partial X_b}{\partial X_1}=0, & \mbox{if $a =1 $}.\end{array} \right. \] 
 If the $n-1$ equations in the first line are satisfied and the first superfield is non-zero $\Phi_1\ne0$,   then the single equation in the second line reduces to $W(X_a)=0$. Since $W(X_a)=0$ is an independent nontrivial equation,  the R-symmetry leaves us with $n$ independent equations for $n-1$ independent variables $X_a$.\footnote{This is not the case for other  global symmetries, because they simultaneously  reduce both the number of independent equations and independent variables.}  In general, this system of equations is over-determined and has no solution which results in spontaneous symmetry breaking.   This is the Nelson-Seiberg argument stating that the existence of an R-symmetry is a necessary condition whereas the spontaneous breakdown of the R-symmetry   is a sufficient condition for spontaneous supersymmetry breaking\cite{Nelson:1993nf}. In the above discussion, we note that the R-symmetry is spontaneously broken because at least one scalar in the first superfield has nontrivial vacuum expectation value due to  $\Phi_1 \ne 0$.  

In the preceding discussion, the verification  of the Nelson-Seiberg argument is based on a supersymmetric theory of chiral superfields. Nevertheless, the supersymmetric Abelian Higgs model we have discussed in  the sections 2 and 3, which is not built  purely from chiral superfields, demonstrates the Nelson-Seiberg argument verbatim. First of all, the Lagrangian of the homogeneous model is R-symmetry invariant with nontrivial vacuum expectation values of the scalar fields, which indicate spontaneously broken R-symmetry. Therefore, in this case the sufficient condition for  spontaneous supersymmetry breaking is met, and supersymmetric solution could not be found. Secondly,  the inhomogeneous mass term explicitly breaks the R-symmetry,  hence, the necessary condition for spontaneous supersymmetry breaking is missing. Therefore,  we could find the inhomogeneous supersymmetric vacuum solution, and this means there is no spontaneous supersymmetry breaking.   Thirdly, the inhomogeneous FI term preserve the R-symmetry, however, the vacuum expectation values of all scalar fields were trivial (R-symmetry is not spontaneously broken), which means we have the necessary but not sufficient condition for the spontaneous supersymmetry breaking. As a result, a supersymmetric vacuum solution could be found also for the model with inhomogeneous FI term. This means there is no spontaneous supersymmetry breaking for the model with  inhomogeneous FI  coefficient as well. 

In order to better understand the effect of inhomogeneity  on the Nelson-Seiberg argument, the inhomogeneous extension of the O'Raifeartaigh model is taken into account~\cite{ORaifeartaigh:1975nky}, which consists purely of chiral superfields and coincides with a spontaneous supersymmetry breaking model in the absence of inhomogeneity. The Lagrangian density of the homogeneous O'Raifeartaigh model is composed of the K\"ahler potential and superpotential of three chiral superfields~\cite{ORaifeartaigh:1975nky}
\begin{align}\label{supfL}
 &K(\Phi_a,\bar\Phi_a)=\sum_{i=1}^3 \bar\Phi_a \Phi_a, \nn\\
 &W(\Phi_a)+ {\rm c.c.}=\Big[ h\Phi_1(\Phi_2^2-\mu^2)+m\Phi_2\Phi_3\Big]+{\rm c.c.},
 \end{align}
 where for the time being we assume all parameters $(h,\mu, m)$ are homogeneous. This preserves R-symmetry and the R-charges of the chiral superfields are given by
 \begin{align}
 R[\Phi_1]=2,\qquad R[\Phi_2]=0,\qquad R[\Phi_3]=2.
\end{align} 
The superpotential is generic because  all the terms allowed by renormalizability and the symmetries of the theory are included. It fulfills the Nelson-Seiberg argument. Indeed, we can write the  Killing spinor equations by setting the variations of the fermionic  fields to zero, 
\begin{align}\label{susysupf-OR}
& \delta\psi_a=i\sqrt2\sigma^\mu\bar\epsilon \partial_\mu\phi_a+\sqrt2\epsilon f_a =0.
\end{align}
Then, after integrating out the auxiliary fields, we can write the BPS equations  as
\begin{align}
\partial_\mu\phi_{a=1,2,3}=0,\quad h(\phi_2^2-\mu^2)=0, \quad 2h\phi_1\phi_2+m\phi_3=0,\quad m\phi_2=0.
\end{align} 
For nonzero mass parameter $\mu\ne0$, there is no solution simultaneously satisfying the second and the fourth equations as long as $m\ne 0$, and thus the supersymmetry is spontaneously broken.  This is often called the F-term breaking~\cite{ORaifeartaigh:1975nky}. 

Now, lets consider the inhomogeneous extension of the O'Raifeartaigh model. If the parameters of the model are  dependent on spatial coordinate, the supersymmetry variation of the superpotential is  not any more  total derivative and is given by
\begin{align}
\int d^2\theta~\delta W+ {\rm c.c.}=\Big(h\partial_\mu[i\bar\epsilon\bar\sigma^\mu C^{1}_{\theta}]-h\mu^2\partial_\mu[i\bar\epsilon\bar\sigma^\mu C^{2}_{\theta}]+m\partial_\mu[i\bar\epsilon\bar\sigma^\mu C^{3}_{\theta}]\Big)+{\rm c.c.},
\end{align} 
where $C^1_{\theta}=\sqrt2(\psi_1\phi_2^2+2\phi_1\phi_2\psi_2),~~C^2_{\theta}=\sqrt2\psi_1$, and $C^3_{\theta}=\sqrt2(\psi_2\phi_3+\phi_2\psi_3)$, respectively,  are the coefficients of the $\theta$-term in the superspace expansions of the three composite/single chiral superfields  $(\Phi_1\Phi_2^2),~\Phi_1$, and $(\Phi_2\Phi_3)$. Then  integration by parts leads to 
\begin{align}
\int d^2\theta~\delta W+{\rm c.c.}=&-i\sqrt2\bar\epsilon\bar\sigma^\mu\Big(\partial_\mu h[\psi_1\phi_2^2+2\phi_1\phi_2\psi_2]-\partial_\mu(h\mu^2) \psi_1+\partial_\mu m[\psi_2\phi_3+\phi_2 \psi_3]\Big)\nn\\
&+{\rm c.c.} +{\rm total~derivative}.
\end{align}
If we assume all the inhomogeneous parameters depend only on a single coordinate $x^1$ and  the projection in \eqref{Sig-epsilon} is applied, then the following expression is obtained
\begin{align}
\int d^2\theta~\delta W+{\rm c.c.} 
=~&i\Big[h'\delta(\phi_1\phi_2^2)-(h\mu^2)'\delta\phi_1+m'\delta(\phi_2\phi_3) \Big] +{\rm c.c.}+{\rm total~derivative}.
\end{align} 
  As usual, the supersymmetry breaking terms can be cancelled if we add the following inhomogeneous deformation terms
\begin{align}
{\cal L}_{\rm Inh}=-i \Big[h'\phi_1\phi_2^2-(h\mu^2)'\phi_1+m'\phi_2\phi_3 \Big]+{\rm c.c.}
\end{align}
In superfield formalism, this can be written as
\begin{align} 
{\cal L}_{\rm Inh} =-i\int d^2\theta~\theta^2 \Big[ h'\Phi_1\Phi_2^2-(h\mu^2)' \Phi_1+m'\Phi_2\Phi_3 \Big]+{\rm c.c.}
\end{align}

 The R-symmetry is explicitly broken by these inhomogeneous deformations, and the model of consideration can not be  a spontaneous supersymmetry breaking model.  
Actually, because of the projection \eqref{Sig-epsilon}, the BPS equations of the inhomogeneous model are modified as  
\begin{align}
&\partial_n \phi_{a=1,2,3}=0,~{\rm for}~n=0,2,3,\nn\\
& i\phi_1'-h(\bar\phi_2^2-\mu^2)=0, \quad i\phi_2'-(2h\bar\phi_1\bar\phi_2+m\bar\phi_3)=0,\quad i\phi_3'-m\bar\phi_2=0.
\end{align} 
These equations support  vacuum solutions. For example, one simple solution is  
\begin{align}
\phi_1(x^1)=i\int_{x_0}^{x^1}h(x)\mu^2(x)dx \quad {\rm and} \quad \phi_2=\phi_3=0.
\end{align}

 In general, any F-term breaking model is necessarily to have a superpotential that involves some linear terms.  Otherwise, the trivial solution $\phi_a=0$ would satisfy the F-term equations and there would be no spontaneous supersymmetry breaking. Therefore, for the spontaneous supersymmetry breaking models, the superpotential should be
\begin{align}
W(\Phi_a)=c_r\Phi_r+{\rm nonlinear~terms},
\end{align}
where $\Phi_r$ are some subset of the  superfields $\Phi_a$. In order for this to be R-symmetric, the R-charges of these selected superfields should be two $R[\Phi_r]=2$. Now, if we consider the inhomogeneous extensions of these models, where the parameters $c_r$ are dependent on one spatial coordinate, then the required inhomogeneous deformations are given by
\begin{align}
~{\cal L}_c=-ic_r'\int d^2\theta~\theta^2\Phi_r+ {\rm c.c.}
\end{align}
Clearly, these inhomogeneous deformations break the R-symmetry, and hence F-term breaking inhomogeneous models can not be obtained due to the Nelson-Seiberg argument.

Consider a theory  containing both chiral and vector superfields in the absence of the FI term. Once all F-terms can be set to zero, which means no F-term breaking, then we can also set all D-terms to zero and there would be no D-term breaking as well. Therefore, if we want to achieve the D-term breaking, we should consider FI term, which is allowed only in Abelian gauge models. On the other hand, in the previous section, we have seen that inhomogeneous extension of a model with the FI term is  not a spontaneous supersymmetry breaking model too. In conclusion, one can say that it is not possible for any inhomogeneous extensions of ${\cal N}=1$ supersymmetric theory to be a spontaneous supersymmetry breaking model.

\section{Conclusion}
 The procedure of obtaining inhomogeneous extension of supersymmetric field theory models is well established. One considers a model with some of its parameters that depend on one or two spatial coordinates, so that  the coordinate dependence breaks supersymmetry explicitly. However, the supersymmetry can be restored partially by imposing some projection condition on the supersymmetry parameters, and then  a required inhomogeneous deformation piece is added to the original Lagrangian.  In this paper, we analyzed the spontaneous supersymmetry breaking phenomena in inhomogeneous field theory models achieved by employing such standard procedure. We presented a detailed analysis of the spontaneous supersymmetry breaking in the inhomogeneous extensions of Abelian Higgs model with the FI term and the O'Raifeartaigh model. We have found that the inhomogeneous deformation pieces either break the R-symmetry or result\ in the models that lack spontaneously broken R-symmetry. Either way, the Nelson-Seiberg argument asserts that these inhomogeneous models can not be  spontaneous supersymmetry breaking models. We have verified this fact by solving the BPS equations and finding the inhomogeneous supersymmetric vacuum solution of zero energy for those inhomogeneous models. Additionally we  have briefly discussed this issue in the inhomogeneous extension of a generic ${\cal N}=1$ model and arrived at the same conclusion. Though we obtained inhomogeneous BPS vacuum solutions, extended objects, e.g. noninteracting topological BPS vortices, of non-zero energy in the constructed inhomogeneous supersymmetric Abelian Higgs models are tackled as an intriguing subject. 
 
 According to the Nelson-Seiberg argument,  inhomogeneous models including spontaneous supersymmetry breaking are possible, if one can find a way to introduce the inhomogeneity in such a way that it preserves the R-symmetry of the Lagrangian but ensures that the vacuum solution breaks the R-symmetry. However,  no such inhomogeneous deformation of ${\cal N}=1$ supersymmetric field theory model is available and we could not  find inhomogeneous spontaneous supersymmetry breaking models in this context. Yet, in extended supersymmetry models, some alternative approaches of introducing inhomogeneity, that are consistent  with the Nelson-Seiberg argument, might exist. In those cases, inhomogeneous supersymmetric field theory models including spontaneous supersymmetry breaking might possibly  be built. We leave this possibility for future investigations.

An interesting byproduct of this work is our assessment about the superfield formalism of inhomogeneous supersymmetric field theory models. We pointed out that, the supersymmetry variation of the Lagrangian density  is determined by  the  coefficients of the terms preceding the highest order terms in the $\theta$, $\bar \theta$ expansions of some gauge-invariant composite/single superfield. If the parameters of the theory are constants, such variation of the Lagrangian density is  a total divergence and the theory is supersymmetric invariant. On the other hand, if the parameters of the theory depend on spatial coordinates,   one can obtain the inhomogeneous deformations  required to make the theory  partially supersymmetric, by applying integration by parts and imposing an appropriate projection on supersymmetry parameters. The superfield form of the resultant inhomogeneous deformation is related to the term that involves the inhomogeneous  parameter in the original Lagrangian, but we should include a factor which is quadratic in $\theta$, $\bar \theta$.   It seems that this procedure is applicable to any supersymmetric model and can be used to formulate the superfield formalism for some well understood inhomogeneous supersymmetric field theory models, such as inhomogeneous ABJM theory \cite{Kim:2018qle,Kim:2019kns} and inhomogeneous mass-deformed SYM theory \cite{Arav:2020obl,Kim:2020jrs}.  We leave this for future investigation as well.

 \section*{Acknowledgments}
 
 We appreciate conversations and discussions with Jeongwon Ho, Chanju Kim,   Sang-A Park, Hanwool Song, Sang-Heon Yi. This work was supported by the
National Research Foundation of Korea(NRF) grant with grant number
NRF-2022R1F1A1073053(Y.K.), NRF-2021R1F1A1062315(D.T.), and RS-2023-00249608, NRF-2019R1A6A1A10073079(O.K.).

\end{document}